\documentclass[prl,pra,twocolumn,superscriptaddress,showpacs,10pt]{revtex4}
\usepackage{graphicx}

\begin{document}

\title{Generating entangled photon pairs from a cavity-QED system}

\author{D.L. Zhou}
\affiliation{School of Physics, Georgia Institute of Technology,
Atlanta, Georgia 30332, USA}
\affiliation{Institute of Theoretical
Physics, The Chinese Academy of Sciences, Beijing 100080, China}

\author{B. Sun}
\affiliation{School of Physics, Georgia Institute of Technology,
Atlanta, Georgia 30332, USA}

\author{C.P. Sun}
\affiliation{Institute of Theoretical Physics, The Chinese Academy
of Sciences, Beijing 100080, China}

\author{L. You}
\affiliation{School of Physics, Georgia Institute of Technology,
Atlanta, Georgia 30332, USA}
\affiliation{Institute of Theoretical
Physics, The Chinese Academy of Sciences, Beijing 100080, China}

\date{\today}

\begin{abstract}
We propose a scheme for the controlled
generation of Einstein-Podosky-Rosen (EPR) entangled photon
pairs from an atom coupled to a high Q optical cavity,
extending the prototype system as a source for
deterministic single photons.
A thorough theoretical analysis confirms the promising operating
conditions of our scheme as afforded by currently available
experimental setups. Our result demonstrates the cavity QED
system as an efficient and effective source for
entangled photon pairs, and shines new light on its
important role in quantum information science.
\end{abstract}


\maketitle

In the Schr\"odinger picture, a quantum
state of a system represents all the knowledge we can obtain.
For a composite system, its wave function or density matrix
describes not only the state of each part, but also the
correlations between the different parts.
The notion of entanglement of a quantum state for
a composite system describes the inseparable correlations
between different parts that are beyond
the classical domain. It has been widely attributed that
entanglement is a valuable resource for quantum computing
and quantum information. Many
current efforts are directed on
the controlled generation and detection of entangled states.

Paradoxically, almost all states of composite systems in
nature are entangled, as a result of interactions among
different system parts. The more useful entangled states,
are those that can be easily and economically
manipulated as in a coupled system of many qubits.
They are the so-called
Einstein-Podosky-Rosen (EPR) state of two qubits \cite{Ein},
Greenberger-Horne-Zeilinger (GHZ) state \cite{Gre}
and W state \cite{Dur} of three qubits,
maximally entangled states \cite{Molmer}
and cluster states \cite{Bri} of many qubits, $\cdots$.
The simplest of them is the EPR state. In a
two qubit (spin-${1}/{2}$) system,
it is commonly denoted as
\begin{eqnarray}
|\text{EPR}\rangle=\frac {1} {\sqrt{2}}
(|\uparrow\rangle_1|\downarrow\rangle_2-|\downarrow\rangle_1|\uparrow\rangle_2),
\label{epr}
\end{eqnarray}
and displays maximal entanglement.
In general, indices 1 and 2 refer to the two qubits,
and in our scheme they refer
to the first and second emitted cavity photons.

Among all physical realizations of qubits, photons are especially
useful as they can be directly used for quantum
communication. The standard process to produce entangled photon
pairs uses nonlinear optical crystals in the so-called parametric
down-conversion process, where a single pump photon spontaneously
decays into an entangled pair composed of a signal and an idler
photon \cite{Kly,Bur}. Despite improvement over the years
with brighter sources for parametric down converted photons,
inherently, the pump photon decay process is stochastic, thus
coincidence counting has to be used. In this paper, we propose a
scheme for deterministically generating entangled photon pairs
from an atom coupled to a high Q optical cavity. Our work is
prompted by the rapid development of a deterministic single
photon source with a trapped ion/atom in a high Q optical
cavity \cite{Park,Law,Duan,Kuhn,Leg,Mac}
and the recent theory for atom-photon entanglement generation and
distribution \cite{Sun}. To our knowledge, this is the
first proposal for deterministically generating EPR photon pairs
from a single atom in a cavity.

Cavity QED is a unique architecture for implementing quantum
computing technology
as it allows for coherent exchange of quantum information
between material qubits (atoms/ions) and photonic qubits.
However, it is a daunting task to reach this goal as this requires
a coherent light-matter coupling at the single photon level,
or the so-called {\it strong coupling} limit.
Several important quantum logic protocols have been developed
within this limit that forms the basis of the
ongoing experimental efforts for quantum computing and communication
with cavity QED systems \cite{cz1,cz2,Kuhn,Mac}.
Another important application of cavity QED is the
``photon gun'' protocol, whereby a single atom leads to
a deterministic cavity photon in the bad cavity limit \cite{Law}.

\begin{figure}[htb]
\includegraphics[width=2.5in]{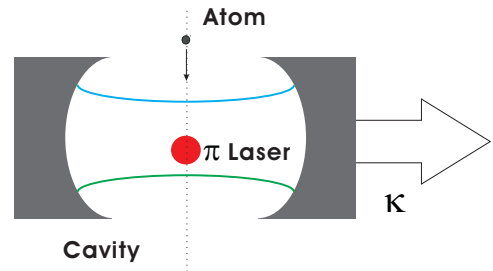}
\caption{(Color online). The illustration of the proposed cavity QED
setup.}
\label{fig1}
\end{figure}

\begin{figure}[htb]
\includegraphics[width=3.25in]{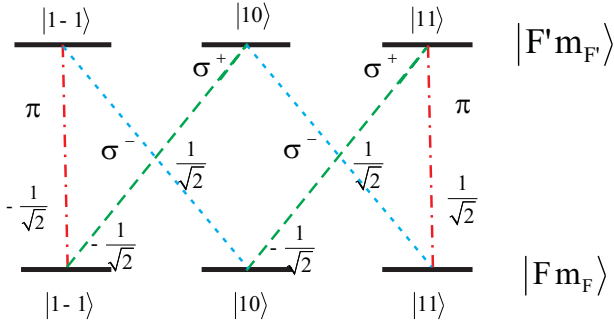}
\caption{(Color online). The proposed coupling scheme: the vertical dot-dashed
lines denote the $\pi$-polarized pump field, while the tilted
lines denote the left (dotted lines for ${\sigma^{-}}$)
and right (dashed lines for ${\sigma^{+}}$) polarized cavity fields.
The various CG coefficients are also indicated.}
\label{fig2}
\end{figure}

The experimental setup for our system is sketched in Fig.
\ref{fig1}. As an atom falls through an optical cavity,
it interacts first with the cavity mode field, then with a
classical pump field $\pi$-polarized with respect to
the cavity axis and propagating along a
perpendicular direction. We utilize
two orthogonal polarizations of the same resonant
cavity mode and assume the atomic levels
to be that of a $F=1\to F'=1$ transition as shown in Fig. \ref{fig2}.
Each Zeeman state is denoted as $|F\,m_F\rangle$,
further simplified as
$|g_{m_F}\rangle=|F\,m_F\rangle$ and
$|e_{m_{F^{\prime}}}\rangle=|F^{\prime}\,m_{F^{\prime}}\rangle$.
Similar model systems were invoked previously in Ref. \cite{Lang},
where Lange {\it et al.} proposed
a scheme for generating GHZ photon multiplets by an
adiabatic passage and in Refs. \cite{Raus,Browne,Marr},
where the entanglement of two modes in one or two
cavities were investigated.

We assume an initially empty cavity and
prepare the atom in state $|e_0\rangle$ before it enters
the cavity mode \cite{Berg}.
The atom's passage through the cavity can now be
divided into two parts. In the first part, the atom only
interact with the cavity mode, and does not enter the area of the
$\pi$-polarized pump field. Thus, the excited atom
emits a first photon, entangled
with the atom in ground states as discussed before \cite{Sun}.
In the second part, the atom is subsequently excited by the pump, and
emits a second photon and swaps its entanglement with the first
photon (already outside the cavity) to the second photon. The whole
process now generates an entangled photon pair. We now analyze
the above protocol including both atomic and cavity decays.
In the interaction picture to the Hamiltonian
\begin{eqnarray}
H_{0}=\hbar\omega_{C}\sum_{m_{F'}}|e_{m_{F'}}\rangle \langle e_{m_{F'}}|
+\hbar\omega_{C}(a_{L}^{\dagger }a_{L}+a_{R}^{\dagger}a_{R})
\end{eqnarray}
of the atom plus the cavity,
our system dynamics is governed by $H=H_{1}+H_{2}$, with
\begin{eqnarray}
H_{1} &=&-\hbar\Delta \sum_{m_{F'}}|e_{m_{F'}}\rangle \langle e_{m_{F'}}|
+{1\over 2}\hbar[\Omega(t)A_{10}^{\dagger }+h.c.],\quad \\
H_{2} &=&{1\over 2}\hbar g(t)( A_{11}^{\dagger }a_{R}+A_{1-1}^{\dagger
}a_{L}+h.c.) ,
\end{eqnarray}%
where the second term of $H_1$ is from the
interaction of the atom with the pump, and
$H_2$ denotes the interaction of the atom with the
left and right circular polarized cavity modes.
$\Delta =\omega_{C}-\omega_{A}$ is the detuning and
for simplicity the pump field is assumed to
be resonant with the cavity.
The atom field coupling coefficients are
\begin{eqnarray}
g({t})&=&-\mathcal{E}(\vec{R})\langle F^{\prime
}||d_{1}||F\rangle,\\
\Omega (t)&=&-E(\vec{R})\langle F^{\prime }||d_{1}||F\rangle,
\end{eqnarray}
where $\langle F^{\prime }||d_{1}||F\rangle$ is the reduced
dipole matrix element according to the Wigner-Eckart theorem,
$\mathcal{E}(\vec{R})$ and $E(\vec{R})$ are respectively the
spatial profiles of the cavity mode and the pump field.
The atomic raising operators are defined as
\begin{eqnarray}
A_{1q}^{\dagger }=\sum_{m_{F},m_{F^{\prime }}}\langle F^{\prime}
m_{F^{\prime }} |F m_{F}1q\rangle |F^{\prime }m_{F^{\prime }}\rangle \langle Fm_{F}|.
\end{eqnarray}

According to the previous prescription, the interaction parameters take the
following time dependence
\begin{eqnarray}
g(t)&=&gh(t)h(T-t),\\
\Omega(t)&=&\Omega h(t-t_1)h(t_2-t),
\end{eqnarray}
with the heaviside step function
$h(t)=0$ for $t<0$ and $h(t)=1$ for $t\ge 0$.
The different times have the following meaning: the atom
enters the cavity at time $0$, arrives at the pump laser
at time $t_1$, leaves the pump at time $t_2$, and finally
exits the cavity at time $T$.

Dissipations are essential to our protocol as they allow
the excited atom to decay and the cavity photons to emit.
Their effects on the dynamics can be included straightforwardly
using a master equation
\begin{eqnarray}
\dot{\rho}=-i[H,\rho ]+D\rho+C\rho,
\label{me}
\end{eqnarray}
where the cavity and atom dissipative terms are
\begin{eqnarray}
D\rho &=&\kappa \sum_{\xi=R,L}(2a_{\xi}\rho
a_{\xi}^{\dagger }-a_{\xi}^{\dagger }a_{\xi}\rho
-\rho a_{\xi}^{\dagger }a_{\xi}), \\
C\rho &=&{{\gamma }\over 2}\ \sum_{q}
(2A_{1q}\rho A_{1q}^{\dagger }-A_{1q}^{\dagger }A_{1q}\rho
-\rho A_{1q}^{\dagger}A_{1q}). \quad
\end{eqnarray}
$2\kappa$ is the one side decay rate of the cavity, while
the other side of the cavity is assumed perfectly reflecting.
$\gamma$ is the decay rate for the atomic excited state $|e_{m_{F'}}\rangle$.
In a more complete model study, we have constructed the wave function for
the system including the free fields outside cavity. We find that our results
are completely verified within the rotating wave approximation \cite{dssy}.

The initial state of our system is now
$|\psi(0)\rangle =|e_{0},0_{L},0_{R}\rangle$, and the rate
of cavity photon emission with polarization
$\xi$ at time $t$ is $p_{\xi}(t)=2\kappa\langle a_{\xi}^{\dagger}(t) a_{\xi}(t)\rangle$.
The probability of cavity photon emission is then
$
P_{\xi }(t) =2\kappa \int_{0}^{t} dt'\,\langle a_{\xi
}^{\dagger }( t') a_{\xi }(t')\rangle.
$
We require that the cavity photon emit quickly as soon as
it is generated, thus the preferred operating condition is
close to the bad cavity limit ($\gamma\ll g^2/\kappa\ll \kappa$)
as for a single photon
source \cite{Law,Duan,Sun}. In fact, we find
it is desirable to operate with
$\kappa \sim g\gg \gamma$,
a compromise of the {\it bad cavity} with
the {\it strong coupling} limit due to the necessity
of coherently pumping the atom to the excited state
for a second photon.

To gain more insight, we describe the
dynamic evolution using the non-Hermitian effective Hamiltonian
\begin{eqnarray}
H_{\rm eff}=H-i\kappa ( a_{L}^{\dagger }a_{L}+a_{R}^{\dagger}a_{R})
-i\frac{\gamma }{2}\sum_{q} |e_q\rangle \langle e_q|.
\label{nh}
\end{eqnarray}
Such an approach is appropriate as re-excitations of the
decayed atom due to emitted photons are negligible,
and re-absorptions of the cavity photons can be neglected due to
their fast decays to outside the cavity.
For $t\in[0,t_1]$, the system state is approximately
\begin{eqnarray}
|\psi (t)\rangle  &=&d_{0}(t) |e_{0},0_{L},0_{R}\rangle +c_{-1}(t)
|g_{-1},0_{L},1_{R}\rangle \nonumber\\
&&+c_{1}(t)
|g_{1},1_{L},0_{R}\rangle .
\end{eqnarray}
The effective Schr\"odinger equation becomes
\begin{eqnarray}
i\frac{d}{dt}\left(
\begin{array}{c}
d_{0}(t)  \\
c_{-1}(t)  \\
c_{1}(t)
\end{array}
\right) =\left(
\begin{array}{ccc}
-\Delta -i\frac{\gamma }{2} & -\frac{g}{\sqrt{2}} & \frac{g}{\sqrt{2}} \\
-\frac{g}{\sqrt{2}} & -i\kappa  & 0 \\
\frac{g}{\sqrt{2}} & 0 & -i\kappa
\end{array}
\right) \left(
\begin{array}{c}
d_{0}(t)  \\
c_{-1}(t)  \\
c_{1}(t)
\end{array}
\right), \nonumber
\end{eqnarray}
which leads to the solution
\begin{eqnarray}
c_{1}(t)=-\frac{ig}{\sqrt{2}}\frac{e^{s_{1}t}-e^{s_{2}t}}{s_{1}-s_{2}},
\end{eqnarray}%
on making use of the property
$c_{1}(t)=-c_{-1}(t)$. $s_{1}$ and $s_{2}$ are the roots of equation
\begin{eqnarray}
2s^{2}+\left( 2\kappa +\gamma -i 2 \Delta \right)
s+\gamma \kappa +2 g^{2}-i 2 \kappa \Delta =0.
\end{eqnarray}
where an identical
coupling is assumed for both polarization modes, thus
$p_R(t)=p_L(t)=2\kappa {| c_1(t)|}^2$.
When the duration before the pump excitation is so long
that the excited atom (in $|e_0\rangle$)
completely decays into ground states and
the cavity photon completely leaks out,
the final state of the atom plus the cavity modes becomes
\begin{eqnarray}
\rho(t_1) &=&\frac{1}{2}|g_{-1},0_L, 0_R\rangle \langle g_{-1},0_L,0_R|\nonumber\\
&&+\frac{1}{2}|g_{1}, 0_L, 0_R\rangle \langle g_{1}, 0_L, 0_R|,
\label{ms}
\end{eqnarray}
as the first photon is traced out after
emitted into modes outside the cavity.

For $t\in(t_1,t_2]$, the initial state (\ref{ms}) is now
completely mixed, so we can evolve its different decompositions
respectively. In the first case for
$|\psi (t_1)\rangle=|g_{1},0_{L},0_{R}\rangle$,
the state can be approximately expanded as
\begin{eqnarray}
|\psi (t) \rangle &=&d_{1}(t)
|e_{1},0_{L},0_{R}\rangle +c_{0}(t) |g_{0},0_{L},1_{R}\rangle \nonumber \\
&& +c_{1}(t) |g_{1},0_{L},0_{R}\rangle,
\end{eqnarray}
and governed by an effective Schr\"odinger equation
\begin{eqnarray}
i\frac{d}{dt}\left(
\begin{array}{c}
d_{1}(t) \\
c_{0}(t) \\
c_{1}(t)
\end{array}
\right) =\left(
\begin{array}{ccc}
-\Delta -i\frac{\gamma }{2} & -\frac{g}{\sqrt{2}} & \frac{\Omega
 }{\sqrt{2}} \\
-\frac{g}{\sqrt{2}} & -i\kappa & 0 \\
\frac{\Omega }{\sqrt{2}} & 0 & 0%
\end{array}
\right) \left(
\begin{array}{c}
d_{1}(t) \\
c_{0}(t) \\
c_{1}(t)
\end{array}
\right). \nonumber
\end{eqnarray}
The solutions for $d_{1}(t)$ and $c_{0}(t)$ are again analytic
and expressed in terms of the roots
$s_{1}$, $s_{2}$, and $s_{3}$ of equation
\begin{eqnarray}
&&2s^{3}+(2\kappa +\gamma -i2\Delta )s^{2}\nonumber\\
&&+(\Omega ^{2}
+\kappa\gamma +g^{2}-i2\Delta \kappa )s+\Omega ^{2}\kappa =0.
\end{eqnarray}

Finally after the atom passes through the $\pi$-polarized
pump field, we find
\begin{eqnarray}
c_{0}(t)=c_{0}\left( t_2\right) \left(
\frac{s_{1}-s_{3}}{s_{1}-s_{2}}
e^{s_{1}(t-t_2)}+\frac{s_{2}-s_{3}}{s_{2}-s_{1}}e^{s_{2}(t-t_2)}\right), \nonumber
\end{eqnarray}
where $s_{1}$ and $s_{2}$ are the roots of equation
\begin{eqnarray}
2s^{2}+\left( 2\kappa +\gamma -i2\Delta \right) s+g^{2}+\kappa
\gamma -i2\kappa \Delta =0,
\end{eqnarray}
and
$s_{3}=-{\gamma }/{2}+i\Delta
-ig{\sqrt{2}d_{1}(t_2)}/[{2c_{0}(t_2)}]$.
The emission rate of a second photon with right circular
polarization during $[t_1,T]$ is therefore
$p_R(t)=\kappa{|c_0(t)|}^2$.

\begin{figure}[htb]
\begin{tabular}{cc}
\includegraphics[width=1.65 in, height=1.30 in]{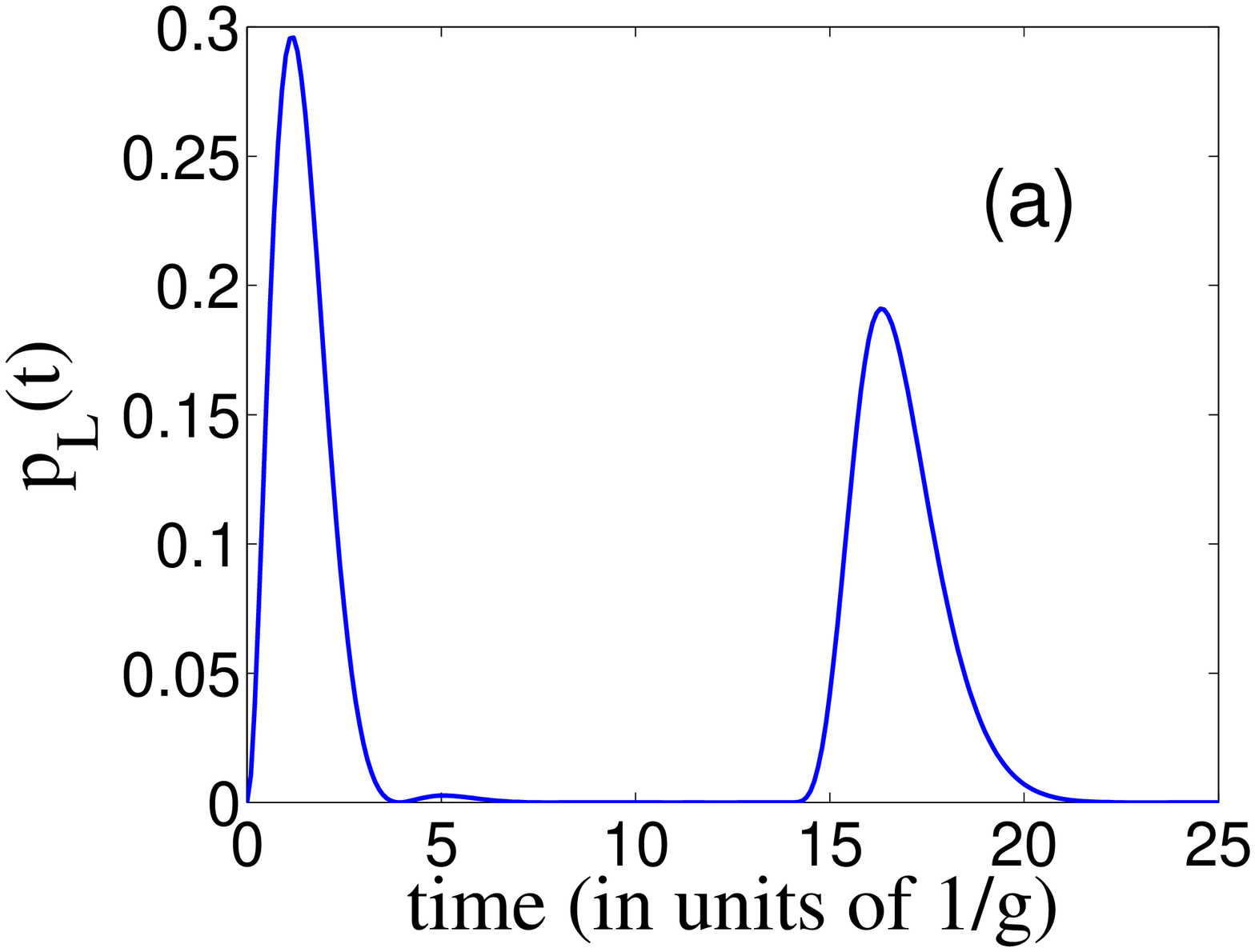}&
\includegraphics[width=1.65 in, height=1.30 in]{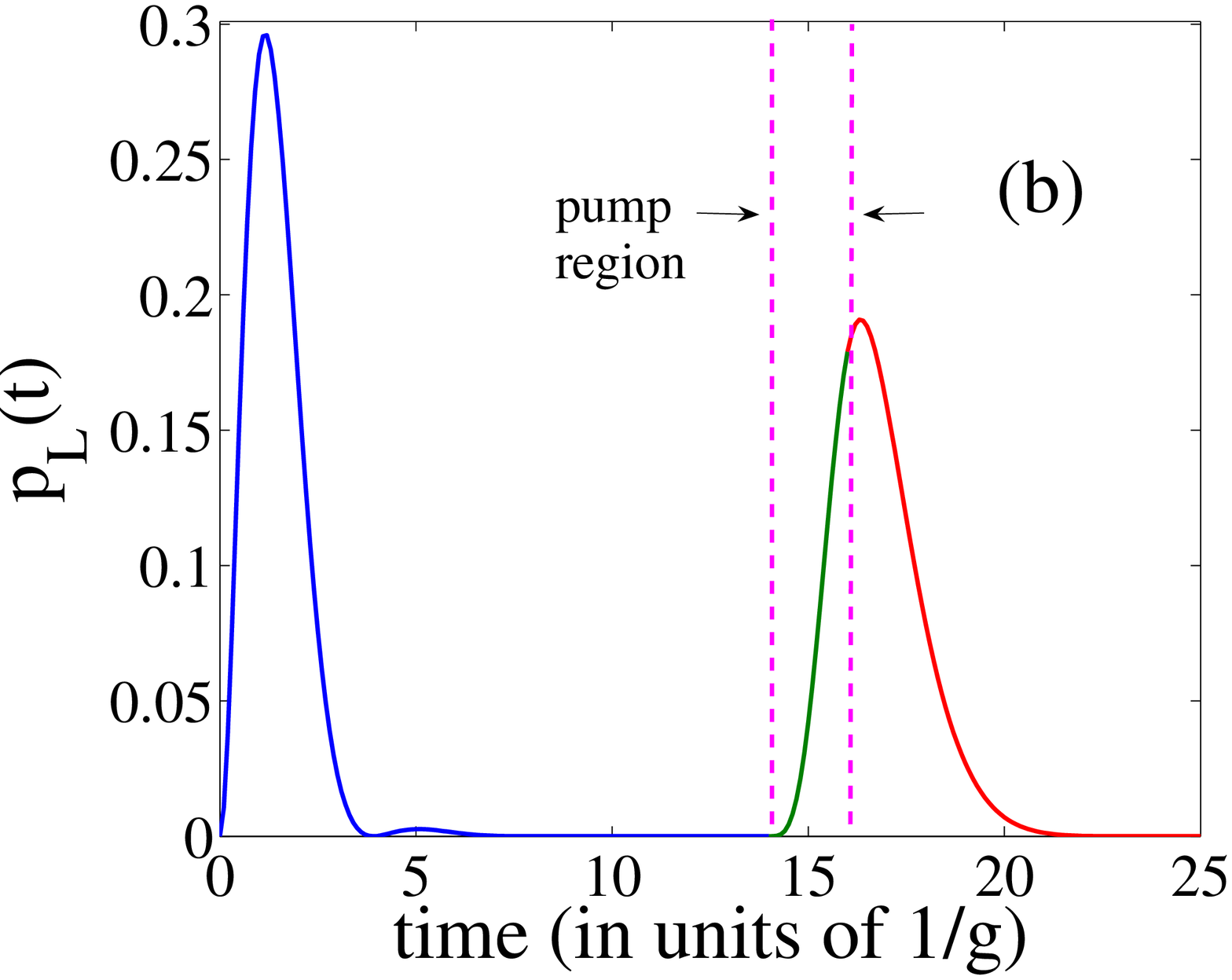}
\end{tabular}
\caption{(Color online). The emission rate of a cavity photon with left circular
polarization at time $t$ based on solving (a) the
master equation (\ref{me}); and (b) the analytic solution
to the non-Hermitian Hamiltonian (\ref{nh}).}
\label{fig3}
\end{figure}

We now compare the above analytical analysis with numerical
solutions from the master equation (\ref{me}) and
the effective non-Hermitian Hamiltonian (\ref{nh}).
For the parameter ranges considered,
the two numerical approaches give the same results as
the analytical one. We find a high fidelity EPR entangled
photon pair of the form (\ref{epr}) ($\uparrow\to 1_L$ and
$\downarrow\to 1_R$) is generated with a high efficiency.
Like nonclassical photon pairs from an atomic
ensemble \cite{kimbles}, these two photons are distinguishable
from their temporal order \cite{simon}, and can be individually
addressed to confirm their EPR correlation.

\begin{figure}[htb]
\begin{tabular}{cc}
\includegraphics[width=1.65 in, height=1.30 in]{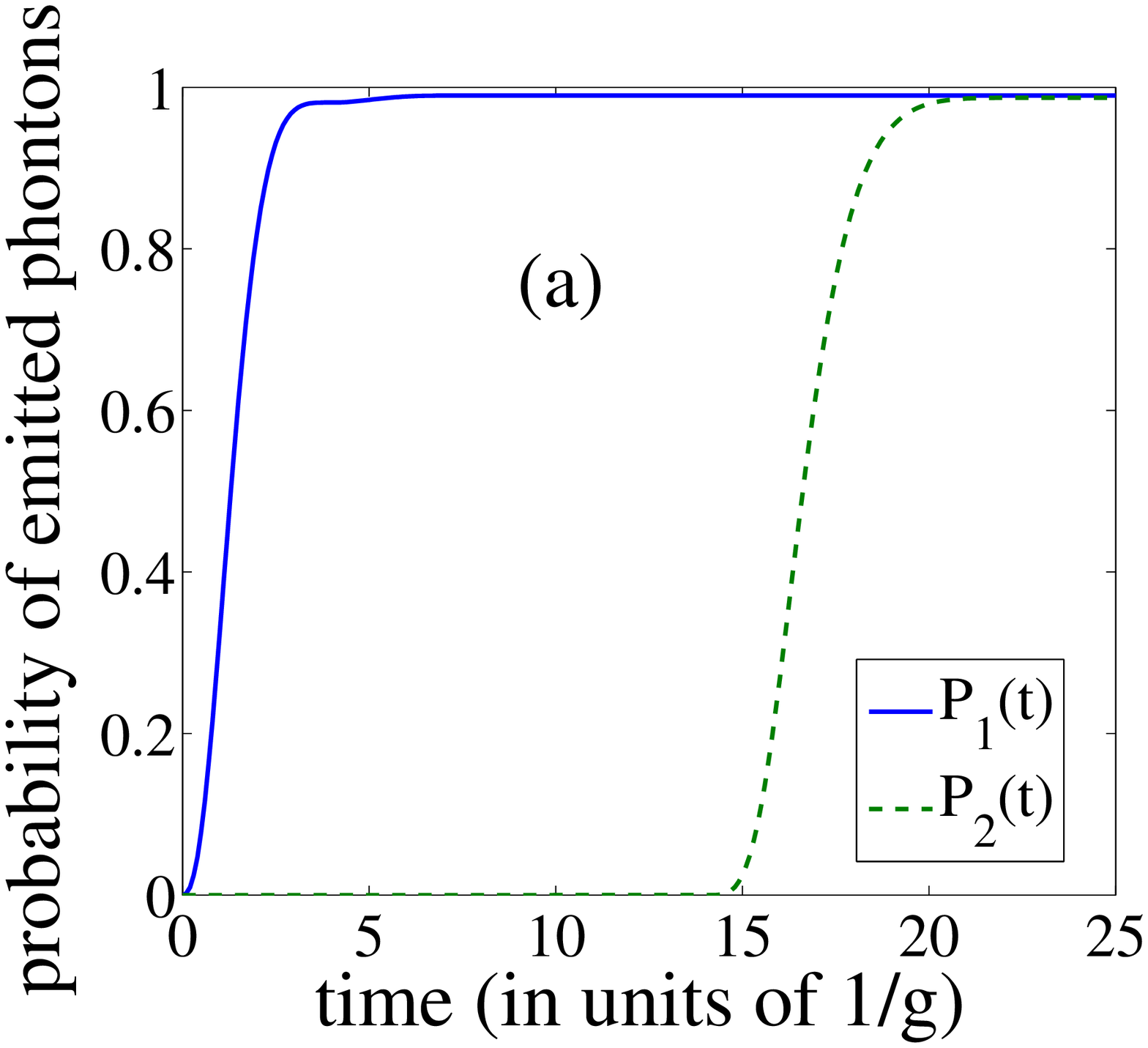}&
\includegraphics[width=1.65 in, height=1.30 in]{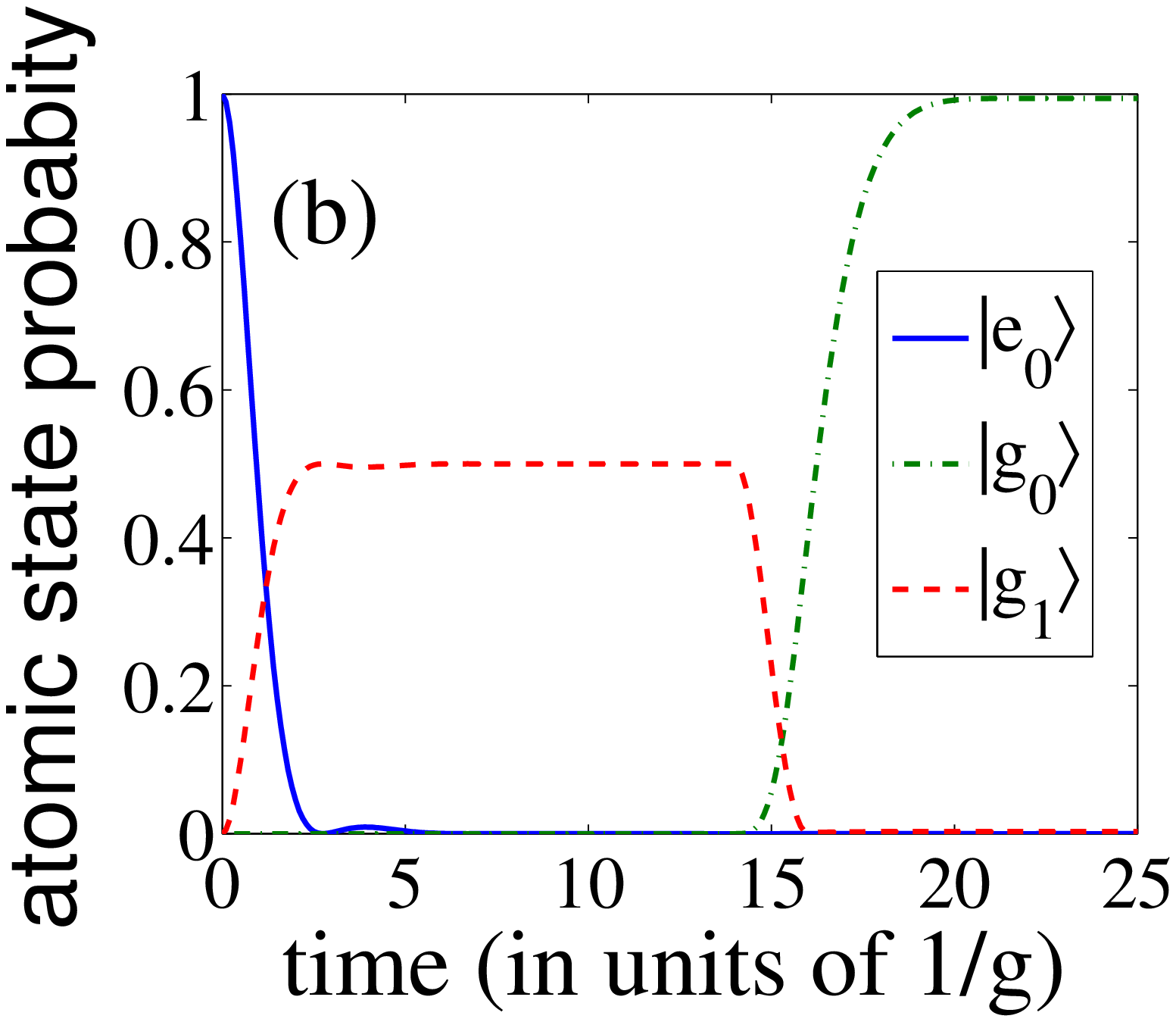}
\end{tabular}
\caption{(Color online). (a) The probability of the first and second photon emission
$P_1(t)$ and $P_2(t)$;
(b) The time dependent atomic state populations.}
\label{fig4}
\end{figure}

In the numerical results shown, we have used
dimensionless parameters
$\Delta=0$, $g=1.0$, $\gamma=0.01$, $\kappa=1.2$, $\Omega=1.2$,
$t_1=14$, $t_2=16$, and $T=25$.
Such a set of parameters
can be realized for corresponding physical parameters of
$\gamma=(2\pi)\,0.2$ (MHz), $g=(2\pi)\,20$ (MHz),
$\kappa=(2\pi)\,24$ (MHz), and $T=200$ (ns) \cite{Sun}.
In Fig. \ref{fig3}, two single photon pulses are seen to
be emitted sequentially from the cavity.
The results from the two numerical approaches
agree well with each other.
In Fig. \ref{fig4}(a), the probability of generating these two single
photons are displayed. We see that they are better than $98\%$.
In Fig. \ref{fig4}(b), the occupations of different atomic states
are shown. Exactly as expected, the first photon is
generated from the decay of state $|e_0\rangle$ to $|g_1\rangle$ or
$|g_{-1}\rangle$; and the second photon is generated from each of
these two states when pumped by the $\pi$-polarized laser.

Based on our extensive simulation with other parameters, we hope
to emphasize three points \cite{Grangier}. First there exists an
optimal $\kappa$, which leads to the fastest photon emission. When
$\kappa\ll g$, oscillations emerge, a signature of the strong
coupling. When $\kappa\gg g$, the probability of emission actually
decays linearly and the emission time becomes longer because of
the increased bandwidth of the cavity, thus correspondingly
reduced emission strength of the atom into the cavity. Second
there is also an optimal time for the atom to pass the pump laser.
When that duration is too short, the atom cannot be completely
emptied from states $|g_{1}\rangle$ and $|g_{-1}\rangle$. However,
if it is too long, oscillations between states $|e_{\pm 1}\rangle$
and $|g_{\pm 1}\rangle$ arise. More conveniently, it will be
desirable to use a trapped atom inside the cavity \cite{Meschede},
and replace the transit time over the pump region with a temporal
pump pulse. Third, the probability of generating the first photon
depends on the atomic decay rate $\gamma$. One solution to
overcome such a dependence is to use an auxiliary starting state
and an additional laser coupled to state $|e_0\rangle$ as in the
single photon source protocol \cite{Law,Duan,Sun}. The
$\pi$-polarization laser then is applied to swap this entanglement
from the atom to the second cavity photon. The entanglement of the
emitted photon pairs can be easily detected from polarization
correlations between the 1st and 2nd photons using a time resolved
detection scheme \cite{simon}.

Finally, we want to emphasize that dissipations are
crucial in our protocol,
not only because they lead to the output of cavity photons.
Without dissipations, the atom cannot enter the state
$|g_0\rangle$, because the quantum amplitudes for the two paths
($|e_0\rangle\rightarrow |g_{-1}\rangle \rightarrow |e_{-1}\rangle
\rightarrow |g_0\rangle$ and $|e_0\rangle\rightarrow |g_1\rangle
\rightarrow |e_1\rangle \rightarrow |g_0\rangle$) interfere
destructively. This balance is broken due to dissipations.

In summary, we have proposed  a simple but efficient scheme to
deterministically generate EPR entangled photon pairs from an atom coupled
to a high Q optical cavity. Our scheme has the potential for realizing
a deterministic source for entangled photon pairs.
The optical nonlinearity of the parametric down conversion
is rather weak and due to many atoms (emitters),
in contrast to the strong resonant interaction of
a single atom as utilized here in a cavity QED.
Thus the down converted photon pairs are inherently
probabilistic. If $p$ (small) is the probability of
a single photon pair, double pairs, triple pairs $\cdots$,
occur with (smaller) probabilities $p^2$, $p^3$ $\cdots$.
In our scheme, however, the photon pair is
deterministic. Each passing atom leads to a single pair, as in a
push-button device.

We thank Dr. M.S. Chapman for helpful discussions.
This work is supported by NSF and NSFC.


\begin{thebibliography}{99}
\bibitem{Ein} A. Einstein, B. Podolsky, and N. Rosen, Phys. Rev.
\textbf{47}, 777 (1935).

\bibitem{Gre} D. M. Greenberger, M. Horne, and A. Zeilinger, in
\textit{Bell's Theorem, Quantum Theory and Conceptions of the
Universe}, edited by M. Kafatos (Kluwer Academic, Dordrecht, The
Netherlands, 1989).

\bibitem{Dur} W. D\"{u}r, G. Vidal, and J. I. Cirac, Phys.
Rev. A \textbf{62} 062314 (2000).

\bibitem{Molmer}K. Molmer, and A. Sorensen, Phys. Rev. Lett. \textbf{82}, 1835 (1999).

\bibitem{Bri} H. J. Briegel and R. Raussendorf, Phys. Rev. Lett. \textbf{86}%
, 910 (2001).

\bibitem{Kly} D. N. Klyshko, Sov. Phys. JETP \textbf{28}, 522
(1969).

\bibitem{Bur} D. C. Burnharm and D. L. Weinberg, Phys. Rev. Lett.
\textbf{25}, 84 (1970).

\bibitem{Park} A. S. Parkins, P. Marte, P. Zoller, and H. J.
Kimble, Phys. Rev. Lett. \textbf{71}, 3095 (1993).

\bibitem{Law} C. K. Law and H. J. Kimble, J. Mod. Opt.
\textbf{44}, 2067 (1997).

\bibitem{Duan}L.-M. Duan, A. Kuzmich, and H. J. Kimble,
Phys. Rev. A {\bf 67}, 032305 (2003).

\bibitem{Kuhn} A. Kuhn, M. Hennrich, and G. Rempe, Phys. Rev.
Lett. \textbf{89}, 067901 (2002); H. J. Kimble,
Phys. Rev. Lett. \textbf{90}, 249801 (2004).

\bibitem{Mac} J. McKeever {\it et al.}, Scince \textbf{303}, 1992 (2004).

\bibitem{Leg} T. Legero, T. Wilk, M. Hennrich, G. Rempe, and A.
Kuhn, Phys. Rev. Lett. \textbf{93}, 070503 (2004).

\bibitem{Sun} B. Sun, M. S. Chapman, and L. You, Phys. Rev. A
\textbf{69}, 042316 (2004).

\bibitem{cz1}T. Pellizzari, S. A. Gardiner, J. I. Cirac, and P. Zoller
Phys. Rev. Lett. {\bf 75}, 3788 (1995).

\bibitem{cz2}J.I. Cirac, P. Zoller, H.J. Kimble, and H. Mabuchi, Phys. Rev.
Lett. {\bf 78}, 3221 (1997).

\bibitem{Lang} W. Lange and H. J. Kimble, Phy. Rev. A \textbf{61},
063817 (2000).

\bibitem{Raus} A. Rauschenbeutel {\it et al.}, Phys. Rev. A
\textbf{64}, 050301(R) (2001).

\bibitem{Browne} D. E. Browne and M. B. Plenio, Phys. Rev. A
\textbf{67}, 012325 (2003).

\bibitem{Marr} C. Marr, A. Beige, and G. Rempe, Phys. Rev. A
\textbf{68}, 033817 (2003).

\bibitem{Berg} K. Bergmann, H. Theuer, and B. W. Shore, Rev. Mod.
Phys. \textbf{70}, 1003 (1998).

\bibitem{dssy} D.L. Zhou, B. Sun, C.P. Sun, and L. You, unpublished.

\bibitem{kimbles}A. Kuzmich {\it et al.}, Nature \textbf{423}, 731 (2003).

\bibitem{simon}C. Simon and Jean-Philippe Poizat,
Phys. Rev. Lett. \textbf{94}, 030502 (2005).

\bibitem{Grangier} P. Grangier, G. Reymond, and N. Schlosser, Fortschr. Phys.
\textbf{48}, 859 (2000).

\bibitem{Meschede}S. Kuhr {\it et al.}, Science \textbf{293}, 278(2001).

\end{thebibliography}
\end{document}